\documentstyle[amsfonts,aps]{revtex}

\begin{document}
\title{Rigged Hilbert Spaces associated with Misra-Prigogine-Courbage Theory of
Irreversibility.}
\author{Adolfo R. Ord\'{o}\~{n}ez}
\address{Facultad de Ciencias Exactas, Ingenier\'{\i}a y Agrimensura.\\
(Universidad Nacional de Rosario)\\
Pellegrini 250, 2000 Rosario, Argentina.\\
I.F.I.R. (Instituto de F\'{\i}sica de Rosario)\\
e-mail : ordoniez@unrctu.edu.ar}
\maketitle

\begin{abstract}
It is proved that, in the Misra-Prigogine-Courbage Theory of Irreversibility
using the Internal Time superoperator, fixing its associated non-unitary
transformation $\Lambda $, amounts to rigging the corresponding
Hilbert-Liouville space. More precisely, it is demonstrated that any $%
\Lambda $ determinates three canonical riggings of the Liouville space $%
{\cal L}$: a first one with a Hilbert space with a norm greater than the
relative one from ${\cal L}$; a second one with a $\sigma $-Hilbertian
space, which is a K\"{o}the space if $\Lambda $ is compact and is a nuclear
space if $\Lambda $ has certain nuclear properties; and finally a third one
with a smaller $\sigma $-Hilbertian space with a still stronger topology
which is nuclear if $\Lambda ^n$ is Hilbert-Schmidt, for some positive
integer $n$. Viceversa: any rigging of this type, originated in a dynamical
system having an Internal Time superoperator, defines a $\Lambda $ in a
canonical way.
\end{abstract}

\section{Introduction.}

The aim of this paper is to demonstrate unsuspected mathematical
implications of the Misra-Prigogine-Courbage theory of irreversibility, one
of the two theories of ''intrinsic irreversibility'' developed by the group
of Brussels (Belgium).

More precisely, it will be proved that {\it MPC-theory is strongly connected
with the theory of Rigged Hilbert Spaces (RHS)}. This is important for two
reasons. First, because it increases the mathematical meaning of the $%
\Lambda $ transformation, and relates it with a well known and successful
subject (specially in Quantum Physics [14][23][24]) as is the Theory of
Riggings. And second, for it makes possible a relation and a comparison with
the other version of irreversibility, namely the ''Rigged Hilbert Space
Extension through the Spectral Decomposition'' [5][6][7][8][9][10].

The Misra-Prigogine-Courbage theory is based on the {\bf Internal Time}
superoperator and its associated the $\Lambda $ transformations
[1][2][3][4]. Let us briefly explain this formalism considering a
hamiltonian system: the motion of the dice when it is thrown. This system is
theoretically deterministic and reversible, but in {\bf reality} it is
impossible to predict the result of one bet, by solving the equations of
motion. It is so because these equations are dynamically {\bf unstable, }%
i.e. any initial condition C is surrounded by many others C' almost
identical to C, but yielding completely different results. Therefore this
kind of dynamics can be considered, ''for all practical purposes'', as a
stochastic process, and solved using the theory of probabilities instead of
newtonian mechanics.

The Brussels School has proved that {\bf any} reversible but unstable enough
dynamic, determines a class of $\Lambda $ transformations. Each member of
the class can be considered as an equivalence ''for all practical purposes''
between the dynamic and a stochastic Markov process, irreversibly convergent
towards an equilibrium density. Moreover, non-isomorphic dynamical processes
are transformed by the $\Lambda $ in non-isomorphic processes [4].

Therefore it is not necessary to use an {\bf arbitrary, observer-dependent,
or extrinsic} ''coarse graining'' {\bf independent of the dynamics }to
transform a reversible evolution into an irreversible one going to
equilibrium. {\it The system itself, if it is sufficiently unstable, defines
its own class of }$\Lambda ${\it \ superoperators, transforming its dynamic
uncertainties (dues to unstability) into probabilistic estimations.}
(Actually, there are systems that also define a class of conditional
expectations that constitute non arbitrary and dynamic-dependent, or
intrinsic ''coarse grainings'' projections, wich also yields an irreversible
Markov process [3])

\section{The $\Lambda (T)$ formalism of the MPC-theory.}

In this section we will introduce a notation that can be used both for
classical and quantum systems

\subsection{The classical case.}

Let us consider an abstract dynamical system [17][18][19]. Let $\Omega $ be
the states space (for example, the phase space), ${\cal B}$ the $\sigma $%
-algebra of measurable sets of $\Omega ,$ and $\mu $ the corresponding
measure (e.g. Liouville measure). Let $S_t:\Omega \rightarrow \Omega $ be
the time-evolution operator on phase space, with $t\in {\Bbb G}$, where $%
{\Bbb G}$ will be ${\Bbb R}$ for the {\bf flows} (i.e. continuous dynamical
systems)and ${\Bbb Z}$ for the {\bf cascades }(discrete dynamical systems). $%
{\frak L}$:$=L^2(\Omega ,{\cal B},\mu )$ will denote the Hilbert space of
the equivalent classes ''a.e.'' (almost everywhere) of measurable functions
of $\Omega $ in ${\Bbb C}$ of square integrable modulus with respect to $\mu
.$ Then, $S_t$ induce an unitary evolution $U_t$ over ${\cal L}$, i.e. a
unitary representation of the group $({\Bbb G},+)$ over ${\cal L}$, defined
as: 
\begin{equation}
(U_t\rho )(\omega )=\rho (S_{-t}(\omega ))\text{ , provided }\rho \in {\frak %
L}\text{, and }\omega \in \Omega  \label{2.1}
\end{equation}
where $U_t$ is known as the Koopman operator{\bf .}

${\cal D}$ will denote the subspace of dimension one generated by the
constant function 1: 
\[
{\cal D}:=\{\alpha 1/\alpha \in {\Bbb C};1:\Omega \rightarrow {\Bbb C}%
1(\omega )=1,\;\text{if}\;\omega \in \Omega \} 
\]
and we will write ${\cal L=D}^{\bot }$. Then: 
\begin{equation}
{\frak L}={\cal D\oplus L}  \label{2.2}
\end{equation}
The positive $\rho \in {\frak L}$ (i.e.$\rho (\omega )\geq 0$ for every $%
\omega \in \Omega $), which are also normal ( in the sense of the $L^1$
norm, i.e. $\int_\Omega \rho (\omega )d\omega =1$), will be the
''probability density functions'' or the Gibbs ''ensembles'' of the system.
If $\mu $ is normalized, in such a way that $\int_\Omega d\mu =\mu (\Omega
)=1,$ then the constant function equal to one is an invariant density under $%
U_t$ as a consequence of eq. (2.1). Furthermore it can be demonstrated that
if the dynamical system is mixing $U_t\rho \rightarrow 1$ in a weak sense.
Therefore 1 is called the equilibrium density and it is symbolized as $%
1=\rho _{eq}$. Also $U_t|_{{\cal L}}\rightarrow 0$ in a weak sense
[17][18][19].

If ${\Bbb G=R},$ and being $U_t$ unitary in ${\frak L}$, there is a
self-adjoint generator $L$ such that: 
\begin{equation}
U_t=e^{-iLt}  \label{2.3}
\end{equation}
If $S_t$ is also a hamiltonian flux, with a hamiltonian function $H$ , then $%
L$ is call the {\bf Liouvillian, }and (2.1) is equivalent to the {\bf %
Liouville differential equation:} 
\begin{equation}
L\rho =i\partial _t\rho  \label{2.4}
\end{equation}
where $L=i\{H,.\}$ and $\{$,$\}$ is the Poisson bracket.

\subsection{The quantum case.}

Non-trivial quantum systems have a continuous spectrum. In this case the
equilibrium state is not an ordinary state but a ''singular diagonal'' state
[15][16][25]. These facts force us to use an extension of the usual quantum
mechanics formalism. Following the line of thought of the cited papers, and
taking into account eq. (2.2) we postulate that our state space is: 
\[
{\frak L}{\cal =D}\oplus {\cal L=}\left\{ \rho =\rho ^d+\rho ^c:\rho ^d\in 
{\cal D}\text{ y }\rho ^c\in {\cal L}\right\} 
\]
Then the $\rho ^{\prime }s$ will evolve under a generalized Liouville
equation [15]. Space ${\cal D}$ contains the information about the {\bf %
probability density} of the states. Space ${\cal L}$ contains the
information about {\bf correlations}, coherent state superposition, and {\bf %
covariance} between observables [16][18]. Then, if the dynamics is mixing we
will have $\rho _t\rightarrow \rho _{eq},$ and $\rho _t^c\rightarrow 0,$ in
a weak sense [17][18].

Let us now consider a quantum system defined in a Hilbert states space $%
{\cal H}$ and its ''complete set of commuting observables''. Let ${\cal D}_{%
{\frak A}}$ be the maximal abelian von Neumann algebra that contains this
set. We will call $I$ to its unit element. If some observables are
essentially selfadjoint unbounded operators, we shall consider the algebra
generated by their spectral projections, which are bounded. Let ${\cal A=L=\{%
}$Hilbert-Schmidt operators over ${\cal H\}}$, with respect to the scalar
product $\langle \rho |\sigma \rangle _{{\cal L}}=Tr(\rho ^{\dagger }\sigma
).$ As it is well known, this space is a Hilbert space. Let us now consider
the {\bf algebra of observables } of the system: ${\frak A=}{\cal D}_{{\frak %
A}}\oplus {\cal A}$, where ${\cal D}_{{\frak A}}$ is the {\bf diagonal part}
of the algebra, and ${\cal A} $ the {\bf non diagonal part }of it{\bf . }Let
us define ${\frak L}$ as the dual space of ${\frak A}$, namely: ${\frak L}=%
{\frak (A)}^{\prime }={\cal D\oplus L}$, where ${\cal D=D}_{{\frak A}%
}^{\prime }$ $=\{\rho _d:\rho _d$ linear and continuous functional over $%
{\cal D}_{{\frak A}}\},$ and where ${\cal L}$ has been identified with $%
{\cal L}^{\prime }.$

The $\rho =\rho ^d+\rho ^c$ are non-negative, in the sense that for any $%
A\in {\frak A}$ , we have: $\rho (A^{\dagger }A)\geq 0,$ where $A^{\dagger
}=(A^d)^{\dagger }+(A^c)^{\dagger }$ and they are normal in the sense that $%
\rho (I)=1$ (as $I\in {\cal D}_{{\frak A}},$ $\rho (I)=\rho ^d(I)).$ These $%
\rho $ will be considered as the possible {\bf states} of the system. The $%
\rho ^d$ which are non-negative as linear functional over the von Neumann
algebra ${\cal D}_{{\frak A}}$ and normal in the sense of $\rho ^d(I)=1,$
will be the {\bf diagonal states, }as are, e.g., the equilibrium states.

Let us observe that if we would take ${\cal A}$ as the set of compact
operators over ${\cal H}$, as in ref. [16], its dual space would be the
space of nuclear operators, which is a subset of the Hilbert-Schmidt
operators [17], namely ${\cal A\subset L}$.

\subsection{The $\Lambda $ transformation.}

Let us consider a continuous linear operator ${\bf \Lambda }{\frak %
:L\rightarrow L}$ , such that:

i) ${\bf \Lambda }$ {\bf preserves probabilities,} in the strong sense that $%
{\bf \Lambda }|_{{\cal D}}=I_{{\cal D}}$ which is the identity in ${\cal D}$%
. I.e., ${\bf \Lambda }=I_{{\cal D}}\oplus \Lambda ,$ where $\Lambda :{\cal %
L\rightarrow L}$ is linear and continuous in the Hilbert space ${\cal L}$.
In particular ${\bf \Lambda }\rho _{eq}=\rho _{eq}.$

ii) ${\bf \Lambda }$ {\bf transforms ensembles into ensembles, }namely ${\bf %
\Lambda }$ preserves the positivity and the normalization. Therefore $%
\Lambda $ must be non negative and symmetric. As the domain of $\Lambda $ is
the whole ${\cal L}$, $\Lambda $ must be self adjoint [20].

iii)${\bf \Lambda }$ {\bf is not a ''coarse-graining''}, namely it doesn't
neglect information as a ''coarse-graining''-projector. It is only a
''change of representation'' that ''reorganizes'', or {\it ''redefines'' the
information content of the densities, in such a way that the resulting
theory is closer to actual experimental possibilities and to physical
reality. }This last requirement is attained by making ${\bf \Lambda }$ {\bf %
an} {\bf injective and dense range application }(states with ''infinite
information content'' are not in the range of ${\bf \Lambda })${\bf . }

In fact, properties i) and ii) above, plus the injectivity, assure that the
range of $\Lambda $ must be either ${\cal L}$ or dense in ${\cal L}$ [20].
If the range of $\Lambda $ is ${\cal L}$, then $\Lambda ^{-1\text{ }}$is
continuous and therefore it is an isomorphism and a homeomorphism, and then $%
\Lambda U_t\Lambda ^{-1}$ is a dynamical system equivalent to $U_t.$ On the
contrary, if the range of $\Lambda $ is dense in ${\cal L}$ then $\Lambda
^{-1}$ is unbounded [20] and this singularity of $\Lambda ^{-1}$ is
essential because it gives new properties to $\Lambda $ that can be
considered as ''catastrophic'' (i.e. with strong ''qualitative changes''
[21]). Precisely the hamiltonian system $U_{t\text{ }}$ is transformed by
the $\Lambda $ into a stochastic process $W_t=\Lambda U_t\Lambda ^{-1}.$ As
now $\Lambda ^{-1}$ is unbounded its domain can be extended beyond the range
of $\Lambda .$

Nevertheless, there is not reason for the positivity of $W_t$ (and therefore
for its markovian character), for any $t\in G$ beyond the range of $\Lambda
. $ So the unboundedness of $\Lambda ^{-1}$ is the crucial ''detail'' that
makes that the $W_t$ do not form a group and breakes the time-symmetry
[1][3].

iv) $W_t=\Lambda U_t\Lambda ^{-1},$ $t\geq 0$ {\bf is the evolution operator
of a strong Markov process}, namely a monotonously convergent process to the
null vector in the Hilbert topology of ${\cal L}$ (and not only in a weak
sense as in the mixing dynamics). I.e.: $||W_t\rho ||_{{\cal L}}\downarrow 0$
if $t\rightarrow \infty ,$ for any $\rho \in Dom(\Lambda ^{-1}).$ This
property is similar to a Markov exact process [18], but in space $L^2$
instead of $L^1.$

Then, $\Lambda ^2$ {\bf is a decreasing Liapounov variable} of the
considered dynamics in the following sense: 
\begin{equation}
\Vert W_t\rho \Vert _{{\cal L}}^2=\langle \Lambda U_t\Lambda ^{-1}\rho \mid
\Lambda U_t\Lambda ^{-1}\rho \rangle _{{\cal L}}=\langle \rho _t\mid \Lambda
^2\rho _t\rangle _{{\cal L}}\downarrow 0  \label{2.5}
\end{equation}
where $\rho _t=U_t\Lambda ^{-1}\rho .$

Accordin to the Brussels group a dynamical system is {\bf intrinsically or
essentially random }if there exists a ${\bf \Lambda :}{\frak L\rightarrow L}$
with the properties above. In order that this happens it is necessary the
mixing character of dynamic, and it is sufficient the existence of an {\bf %
age or internal time operator} $T$ [22][1][3][25]. For the flows $T$ is a
kind of ''time-position operator'', similar to the ''space-position''
operator $Q$ of quantum mechanics, but it acts in space ${\cal L}$ instead
of space ${\cal H}$ [2].

The Liouville operator $L$ is the ''canonical conjugate momentum'' of $T$ : 
\begin{equation}
\lbrack T,L]=i\Longleftrightarrow U_t^{\dagger }TU_t=T+t,\;t\in {\Bbb R}
\label{2.6}
\end{equation}
This last equation between the ''internal time'' and the ''external time'' $%
t $, which can also be used for cascades, can be considered as a general
definition of $T:$%
\begin{equation}
U_t^{\dagger }TU_t=T+t,\;t\in {\Bbb G}  \label{2.7}
\end{equation}
The construction of operator $T$ is completely similar to that of operator $%
Q.$ $T$ exist iff the unitary representation $U_t$ of (${\Bbb G}$,${\Bbb +)}$
in ${\cal L}$ is {\bf imprimitive} with respect to ${\Bbb G}$ . This means
that there is a spectral measure $E$, defined over a $\sigma $-algebra $%
{\cal B}$ of ${\Bbb G}$, and whose values are orthogonal projectors of $%
{\cal L}$ [20][23], such that: 
\begin{equation}
U_t^{\dagger }E(\Delta )U_t=E(\Delta +t),\text{ for }t\in {\Bbb G},\text{
and }\Delta \in {\cal B}  \label{2.8}
\end{equation}
In such a case: 
\begin{equation}
T=\int\limits_{{\Bbb G}}s\text{ }dE  \label{2.9}
\end{equation}

From eq. (2.8), for all $\rho \in {\cal L}$, the numerical measure $\Delta
\rightarrow \langle \rho |E(\Delta )\rho \rangle _{{\cal L}}$, is
translational invariant. Then, if ${\Bbb G=R},$ it is equivalent to the
Lebesgue measure [13]. In other words, for flows, the spectrum of $L$ must
be absolutely continuous and uniform [19][6]. This condition is fulfilled
for classical and quantum K-flows [3][25][26]. Going back to the general
case, any ${\cal B}$-measurable function $\lambda :{\Bbb G}\rightarrow
[0,1], $ such that:

i) $\lambda $ is decreasing, i.e.: $r<s\Rightarrow \lambda (r)\geq \lambda
(s).$

ii)$\lambda (t)\rightarrow 1$ if $t\rightarrow -\infty $ and $\lambda
(t)\rightarrow 0$ if $t\rightarrow \infty .$

iii) If $t\geq 0:\frac{\lambda (s+t)}{\lambda (s)}\downarrow 0,$ i.e.: $%
r<s\Longrightarrow \frac{\lambda (r+t)}{\lambda (r)}\geq \frac{\lambda (s+t)%
}{\lambda (s)}$ and $\frac{\lambda (s+t)}{\lambda (s)}\rightarrow 0$ if $%
s\rightarrow \infty .$

defines a ${\bf \Lambda }$ as: 
\begin{equation}
\Lambda =\lambda (T)=\int\limits_{{\Bbb G}}\lambda (s)\text{ }dE
\label{2.10}
\end{equation}
\begin{equation}
{\bf \Lambda }=I_{{\cal D}}\oplus \Lambda  \label{2.11}
\end{equation}
Since operator $T$ fulfills eq. (2.7), we have:

\begin{eqnarray}
U_t^{\dagger }\lambda (T)U_t &=&\lambda (T+t),\text{for }t\in {\Bbb G} \\
U_t^{\dagger }\lambda ^2(T)U_t &=&\lambda ^2(T+t),\text{for }t\in {\Bbb G}
\end{eqnarray}
where $\lambda ^2(T)=\Lambda ^2$ is the decreasing Liapounov variable.

\section{The rigged Hilbert spaces.}

Let ${\cal L}$ be a separable Hilbert space (e.g. a Liouville space, but
here it will be considered in a general sense). Let $\Psi $ be a proper
vector subspace of ${\cal L}$. Let us suppose that in $\Psi $ it is defined
a countable family of Hilbert norms $\{||.||_n\}_{n\in N}$ , where $%
||.||_n=\langle .|.\rangle _n^{\frac 12},$ $N\subset {\Bbb Q}^{+}$ (the set
of rational non negative numbers, therefore the set $N$ is countable), such
that:

(i) $n_1\leq n_2\Longrightarrow ||.||_{n_1}\leq ||.||_{n_2}$ and both norms
are {\bf compatible}, meaning that if $\{\rho _n\}$ is a Cauchy sequence in
both norms, and if $||\rho _m||_{n_1}\rightarrow 0$ then $||\rho
_m||_{n_2}\rightarrow 0.$

(ii) $N$ has a minimum element, that wil be assumed to be zero (for
simplicity), $||.||_0=||.||_{{\cal L}}$, and $\Psi $ is dense in ${\cal L}$
(meaning that the completion of $(\Psi ,||.||_0)$ is ${\cal L}$)

In such a case, the completion of $\Psi $ with $||.||_{n_i}$ will be denoted 
$\Phi _{n_i}$, whose elements are the equivalence classes $[\{\rho
_m\}]_{n_i}$ of the Cauchy sequences $\{\rho _m\}$ of $\Psi ,$ where: 
\[
\{\rho _m\}\sim \{\sigma _m\}\Leftrightarrow \left\| \rho _m-\sigma
_m\right\| _{n_i}\rightarrow 0\text{ if }m\rightarrow \infty 
\]
\[
\Psi \subset \Phi _{n_2}\subset \Phi _{n_1}\subset \Phi _0={\cal L} 
\]
condition (i) assures the injectivity and continuity of the canonical
applications $i_{n_2,n_1}:\Phi _{n_2}\rightarrow \Phi _{n_1}$ defined by $%
[\{\rho _m\}]_{n_2}\rightarrow [\{\rho _m\}]_{n_1},$ and condition (ii) that
they have a dense range.

Let us now consider the local convex topology defined by the family of norm
on $\Psi .$ In other words the topology such that: 
\[
\{\rho _m\}\rightarrow \rho \Longleftrightarrow \text{for every }n\in
N:\left\| \rho _m-\rho \right\| _n\rightarrow 0 
\]
There are three possibilities:

\noindent {\bf A) }$N${\bf \ has a maximum \~{n}.}

\noindent In this case, from condition (i) we have: 
\[
\{\rho _m\}\rightarrow \rho \Longleftrightarrow \left\| \rho _m-\rho
\right\| _{\tilde{n}}\rightarrow 0 
\]
But this is the norm topology $||.||_{\tilde{n}}.$ Then, by completion of $%
(\Psi ,||.||_{\tilde{n}})$ we get a Hilbert space $\Phi _H,$ such that for
every $n\in N:\Phi _H\subset \Phi _n.$

\noindent {\bf B) }$N$ {\bf has a supreme \~{n}, but \~{n} }$\notin N$.

\noindent In this case we do not get a Hilbert space, but a $\sigma -${\bf %
Hilbertian space }[12], that we shall call $\Phi _K$ (because, as we shall
see later, this is a K\"{o}the space, if $\Lambda $ is compact). In
particular, if $N$ has a maximum $\tilde{n},$ but we only consider the
family of norms $\{||.||_n\}_{n\in N-\{\tilde{n}\}}$, we get a space $\Phi
_K=\cap \{\Phi _n:n\in N-\{\tilde{n}\}\}$ such that $\Phi _H\subset \Phi _K$.

\noindent {\bf C) }$N${\bf \ is not bounded from above.}

\noindent In this case we get a smallest $\sigma -${\bf Hilbertian space}
with a stronger topology than $\Phi _K$ (and therefore easier to transform
in a {\bf nuclear} topology, by endowing $\Lambda $ with more properties$).$
We shall call this space $\Phi .$ Precisely, there is a sequence of subsets
of $N$, $\{N_p\}$, such that: $N_1\subset N_2\subset ...$, and $%
\bigcup_pN_p=N.$ In this way we can obtain a sequence of spaces $\Phi _{H_p}$
as in paragraph (A). Now, condition (i) assures that the canonical mappings $%
i_{H_p}:\Phi _{H_p}$ $\rightarrow {\cal L}$, defined as the $i_{n_2,n_1}$
when $n_2=1,$ $n_1=0,$ and the mapping $i:\Phi \rightarrow {\cal L}$,
defined as: $i(\widetilde{\rho })=i_{H_p}(\widetilde{\rho }),$ for every $%
\widetilde{\rho }\in \Phi $ and every $p$, are all of them injectives and
continuous, and condition (ii) assures that they have a dense range.

In any of these cases it is usual to say that {\bf we have rigged the
Hilbert space }${\cal L}$ with another Hilbert space $\Phi _H$ or with a $%
\sigma -$Hilbertian space, either $\Phi _K$ or $\Phi .$ Really we must also
consider the corresponding antidual spaces (of continuous antilinear
functions) that we shall call $\Phi _H^{\times },$ $\Phi _K^{\times },$ $%
\Phi ^{\times },$ ${\cal L}^{\times }={\cal L}$. In fact, as the topologies
of $\Phi _H,$ $\Phi _K,$ and $\Phi $ are stronger than that of ${\cal L}$,
they make possible the existence of larger sets of continuous antilinear
functionals. Therefore we have: 
\[
\Phi _H\subset {\cal L}\subset \Phi _H^{\times }\text{ };\text{ }\Phi
_K\subset {\cal L}\subset \Phi _K^{\times }\text{ };\text{ }\Phi \subset 
{\cal L}\subset \Phi ^{\times } 
\]
where the corresponding inclusions are continuous and their images are dense.

Let us consider a rigging of type (A). Let ${\cal R}:\Phi _H^{\times
}\rightarrow \Phi _H$ be the Riesz representation: to every antilinear
continuous functional $F$ it associates the vector $\rho _F$ such that: 
\begin{equation}
\left\langle \sigma \mid \rho _F\right\rangle _{\Phi _H}=F(\sigma )\text{ ,
for every }\sigma \in \Phi _H  \label{3.1}
\end{equation}
It is known that ${\cal R}$ is an isometric isomorphism. Nevertheless, $\Phi
_H\neq \Phi _H^{\times }$ if we consider these spaces just like sets. Then: 
\[
\Phi _H\subset {\cal L}\subset \Phi _H^{\times } 
\]
It is easy to see that ${\cal R}|_{{\cal L}}=R$ is a non negative operator.
In fact, from Riesz representation of ${\cal L}^{\times }$ in ${\cal L}$,
all $\rho \in {\cal L}$ can be considered as an antilinear continuous
functional on ${\cal L}$, 
\begin{equation}
\sigma \mapsto \left\langle \sigma \mid \rho \right\rangle _{{\cal L}}
\label{3.2}
\end{equation}
On the other hand, as ${\cal L}\subset \Phi _H^{\times }$ the same $\rho $
can be thought as a functional: 
\begin{equation}
\sigma \mapsto \left\langle \sigma \mid R(\rho )\right\rangle _{\Phi _H}
\label{3.3}
\end{equation}
Then: 
\begin{equation}
\left\langle \sigma \mid \rho \right\rangle _{{\cal L}}=\left\langle \sigma
\mid R(\rho )\right\rangle _{\Phi _H}  \label{3.4}
\end{equation}

\noindent If, in particular, $\sigma =R(\rho ),$ then: 
\begin{equation}
\left\langle R(\rho )\mid \rho \right\rangle _{{\cal L}}=\left\langle R(\rho
)\mid R(\rho )\right\rangle _{\Phi _H}\geq 0  \label{3.5}
\end{equation}
Therefore $R$ is non negative and thus it has a square root $J=R^{\frac 12},$
which also is non negative, continuous and self-adjoint in ${\cal L},$
injective and with dense range. Furthermore it is proven in ref. [11] that: 
\begin{equation}
\left\langle J\sigma \mid J\rho \right\rangle _{\Phi _H}=\left\langle \sigma
\mid \rho \right\rangle _{{\cal L}}  \label{3.6}
\end{equation}

\noindent in such a way that $J$ turns out to be an isometry.

Viceversa., if we have an operator $J:{\cal L\rightarrow L}$ with the same
properties as above, then the relation: 
\begin{equation}
\left\langle J\sigma \mid J\rho \right\rangle _\Psi =\left\langle \sigma
\mid \rho \right\rangle _{{\cal L}}  \label{3.7}
\end{equation}
defines a scalar product on $\Psi =Ran(J),$ whose completion is a Hilbert
space $\Phi _H$ which riggs ${\cal L}$ in a canonical way. Let us remark
that giving a rigging type (A) is equivalent to giving an operator $J$ with
the properties listed above, that we shall call the {\bf associated operator 
}to the rigging. Let us also observe that, as the range of $J$ is dense, $%
J^{-1}$ is unbounded.

Furthermore, an operator $J$ defines a canonical rigging of type (B) and
another one of type (C). In fact, let $N_B=\{n=p/p+1:p=0,1,2,...\}$ and let $%
N_C=\{0,1,2,...\}.$ Let us also define on $\Psi $ the Hilbert norms: 
\begin{equation}
\Vert J\rho \Vert _n^2=\left\langle J^n\rho \mid J^n\rho \right\rangle
_n=\left\langle \rho \mid \rho \right\rangle _{{\cal L}}  \label{3.8}
\end{equation}
where $n\in N_{B\text{ }}$ in the first case, and $n\in N_C$ in the second
one. (If we had chosen as $N_B$ any other subset of rational, non-negative
numbers with a supreme equal to 1, we would have had another equivalent
sequence of norms, yielding the same $\Phi _K.$ If we had chosen as $N_C$
another growing sequence of rational numbers we would have had the same
space $\Phi ).$

Even if $J:{\cal L\rightarrow }\Phi _H$ is an isometry, this would not be
the case for $J,$ considered as an operator $J:{\cal L\rightarrow L}$. In
fact, the difference between the second operator and the first one, is that
the latter establishes the ''deformation power'' of the former. For
instance, the first one maps the unit sphere $S$ of ${\cal L}$ in the unit
sphere $S_H$ of $\Phi _{H.}$ On the other hand, if the second one is {\bf %
bounded or continuous,} $S_H$ is only a bounded set of ${\cal L}$. But if it
is compact [11][13], $S_H$ will be an ellipsoid whose semiaxis go to zero,
i.e. the action of $J$ is more ''drastic''. If it is {\bf Hilbert-Schmidt or
nuclear} [11][13][17], $S_H$ will be an ellipsoid with semiaxis going to
zero in $l^2$ or in $l^1$, respectively.

\noindent {\bf Example 1.}

\noindent Let us suppose that $J:{\cal L\rightarrow L}$ is a compact
operator. Then we will show that $\Phi _K$ is a K\"{o}the space [11][13]. As 
$J$ is compact, there is an orthonormal basis $\{\rho _k\}$ of ${\cal L}$
and a sequence of numbers $\lambda _k\geq 0,$ $\lambda _k\downarrow 0$, such
that: 
\[
\text{for every }\sigma \in {\cal L}\text{, it is: }\sigma
=\sum_{k=1}^\infty a_k\rho _k\text{ , and }J\sigma =\sum_{k=1}^\infty
\lambda _k\left\langle \sigma \mid \rho _k\right\rangle _{{\cal L}}\rho _k 
\]

\noindent Then for any $n\in N_B$, we have:

\begin{eqnarray*}
\left\langle J^n\sigma \mid J^n\rho \right\rangle _n &=&\left\langle \sigma
\mid \rho \right\rangle _{{\cal L}}=\sum_{k=1}^\infty
a_k^{*}b_k=\sum_{k=1}^\infty \left\langle \sigma \mid \rho _k\right\rangle _{%
{\cal L}}^{*}\left\langle \rho \mid \rho _k\right\rangle _{{\cal L}}= \\
&=&\sum_{k=1}^\infty \left\langle J^n\sigma \mid \rho _k\right\rangle _{%
{\cal L}}^{*}\left\langle J^n\rho \mid \rho _k\right\rangle _{{\cal L}%
}\lambda _k^{-2n}
\end{eqnarray*}
In other words: 
\[
\Phi _n=l^2(\lambda _k^{-n})=\left\{ \left\{ a_k\right\} /\sum_{k=1}^\infty
\left| a_k\right| ^2\lambda _k^{-2n}<\infty \right\} 
\]

\noindent and $\Phi _K=\bigcap_{n\in N_B}\Phi _n$ , endowed with the
sequence of Hilbert norms of all these spaces $\Phi _n,$ is by definition, a
K\"{o}the-Toeplitz space.

\noindent Furthermore if $J$ is not only compact but also satisfies the
condition: 
\begin{equation}
\overline{l\acute{\imath}m}\frac{\lambda _{k+1}}{\lambda _k}<1  \label{3.9}
\end{equation}
(which in particular, using the quotient theorem for series, implies that $%
\sum_{k=1}^\infty \lambda _k<\infty $, and therefore that $J$ is nuclear)
then it is attained a necessary and sufficient condition for $\Phi _K$ being
a K\"{o}the nuclear space, namely: for every $n_1\in N_B$ there exists a $%
n_2\in N_B,$ $0\leq n_1<n_2<1,$ such that: 
\begin{equation}
\sum_{k=1}^\infty \frac{\lambda _k^{-2n_1}}{\lambda _k^{-2n_2}}<\infty
\label{3.10}
\end{equation}
In fact, using the same quotient theorem, but now in the serie (3.10), we
have: 
\begin{equation}
\frac{\frac{\lambda _{k+1}^{-2n_1}}{\lambda _{k+1}^{-2n_2}}}{\frac{\lambda
_k^{-2n_1}}{\lambda _k^{-2n_2}}}=\left( \frac{\lambda _{k+1}}{\lambda _k}%
\right) ^{2(n_2-n_1)}  \label{3.11}
\end{equation}
But from eq. (3.9) and considering that $n_2-n_1>0$ we have: 
\begin{equation}
\overline{l\acute{\imath}m}\left( \frac{\lambda _{k+1}}{\lambda _k}\right)
^{2(n_2-n_1)}<1  \label{3.12}
\end{equation}

\noindent {\bf Example 2.}

\noindent Let us now consider the most typical quantum mechanics rigging, as
explained in [14]. The physical quantum system is represented by an algebra
of observables, acting on a vector space with an inner product $\langle
.|.\rangle .$ $R_n$ will denote the eigenspace of the eigenvalue $n$ of the
hamiltonian of the system. $N$ will be the ''number of modes'' operator of
the system$.$ If $N_C=\{0,1,2,...\},$ let $\Psi =\bigoplus\limits_{n\in
N_C}R_n,$ be the set of all those states that are {\bf finite} linear
combinations of the energy eigenstates of the system. For each $n\in N_C,$
let us define a scalar product $\langle .|.\rangle _n$ on $\Psi $ as: 
\begin{equation}
\left\langle \phi \mid \psi \right\rangle _n=\left\langle \phi \mid
(N+I)^n\psi \right\rangle \text{ for every pair }\phi \text{ and }\psi \text{
of }\Psi  \label{3.13}
\end{equation}
or which is equivalent, if $J=(N+I)^{-\frac 12},\phi =J^n\sigma ,\psi
=J^n\rho :$%
\begin{equation}
\left\langle J^n\sigma \mid J^n\rho \right\rangle _n=\left\langle
(N+I)^{-n/2}\sigma \mid (N+I)^{-n/2}\rho \right\rangle _n=\left\langle
\sigma \mid \rho \right\rangle  \label{3.14}
\end{equation}
In this way we have defined a family of Hilbert norms: 
\begin{equation}
\left\| \psi \right\| _n=\sqrt{\left\langle \psi \mid \psi \right\rangle _n}
\label{3.15}
\end{equation}
and the corresponding rigging of the Hilbert space ${\cal H}$ obtained by
the completion of $\Psi $ with the norm (3.15) for $n=0$. In this case, the
associated mapping $J$ , has the spectrum: 
\begin{equation}
\{\lambda _k:k=1,2,...\}=\left\{ \frac 1{\sqrt{k+1}}\text{ : }%
k=1,2,..\right\}  \label{3.16}
\end{equation}
Then, $J^{-1}$ has spectrum $\{\surd \overline{k+1}:k=1,2,...\}$, which is
an unbounded set, and therefore $J^{-1}$ is an unbounded operator.
Furthermore, it is obvious that, even if $J$ is neither a nuclear operator
(since $\sum_{k=1}^\infty \lambda _k=\infty )$ nor a Hilbert-Schmidt one
(since $\sum_{k=1}^\infty $ $\lambda _k^2=\infty ),$ the powers of $J$ are
nuclear operators. In fact, as $J$ $^2$ is a Hilbert-Schmidt operator, it
follows that $J^n$ is nuclear for every $n\geq 4.$ That's why, in this case,
we have rigged ${\cal H}$ with a nuclear space. It's important to remark
that this is a rigging of ${\cal H}$, not of ${\cal L}$, and that this
system is dynamically stable. So, this $J$ is not a $\Lambda $ in the sense
of the Brussels group.

Finally let us observe that there exists riggings of more general types
[11], i.e. using nonmetrizable spaces. But they are not relevant for this
paper.

\section{Equivalence $\Lambda (T)$-coherent rigging.}

From all we have said it is obvious that, given a dynamical system with an
internal time operator $T$ and a ${\bf \Lambda }=I_D\oplus \Lambda $ such
that $\Lambda =\lambda (T),$ then there is a canonic operator $J$ defined as 
$J=\Lambda $, endowed with the necessary properties to define a rigging of $%
{\cal L}$ of each one of the three types (A), (B), and (C). This rigging is
deeply related with the dynamics, since it is defined through a $\Lambda $,
and eq. (2.8) is valid. This means that {\it a relation exists between that
family of growing norms, defined by }$\Lambda ${\it \ via eqs. (3.6) or
(3.8) and the time evolution of the dynamics. In fact, both }$\Lambda ${\it %
\ and the Liapunov variable }$\Lambda ^2=R${\it , are ''decreasing
functions'' of }$T,${\it \ in the sense that they are respectively equal to }%
$\lambda (T)${\it \ and }$\lambda ^2(T),${\it \ being }$\lambda
(t)\downarrow 0${\it \ for }$t\rightarrow \infty .$ In these cases we will
say that the rigging is {\bf coherent} with the dynamics.

Since a $\Lambda $ defines a type (A) rigging, we have a base to say that
the stochastic process whose semigroup of contracting operators is $%
W_t=\Lambda U_t\Lambda ^{-1}$ in ${\cal L}$ is, in some sense,
''equivalent'' (for $t>0$) to the dynamical system whose group of unitary
evolution is $U_t.$ In fact, as we have said $\Lambda :{\cal L\rightarrow }%
\Phi _H$ is an isometry and therefore, in $\Phi _H$, the above process is
the {\bf isometric image}, for $t\geq $ $0,$ of the $U_t$ dynamics.
Something very similar, but not so ''perfect'', happens if we use the type
(B) rigging. In this case we have a sequence of isometries $\{\Lambda ^n:%
{\cal L}\rightarrow \Phi _n\}$ with $n\in N_B,$ being $\Lambda ^n\rightarrow
\Lambda $ if $n\rightarrow 1$ ($n=p/p+1$ if $p\rightarrow \infty ).$ Thus $%
W_t$ turns out to be the limit of a sequence of isometric images of $U_t.$

Let us now consider any rigging of ${\cal L},$ which is coherent with a
dynamic that has an internal time $T$. This rigging may be of any of the
three types (A), (B), or (C), and it must be defined by a unique $J=\Lambda
=\lambda (T),$ with $\lambda :{\Bbb R}\rightarrow [0,1]$ endowed with the
properties listed above the eq. (2.6). Then let us define : ${\bf \Lambda }%
=I_{{\cal D}}\oplus \Lambda .$ In this way we have all the properties of $%
{\bf \Lambda }$ with the exception of the normalization and the monotonous
convergence of $||W_t\rho ||$ to zero.

The normalization turns out to be trivial in the quantum case, since it is
defined by the diagonal part: if $\rho $ is normal then: 
\begin{equation}
\left( {\bf \Lambda }\rho \right) (I)=(I_{{\cal D}}\rho ^d)(I)=\rho ^d(I)=1
\label{4.1}
\end{equation}

In the classical case, if $\rho $ is normal, namely if $\int_\Omega \rho
d\mu =1,$ then: 
\begin{equation}
\int_\Omega {\bf \Lambda }(\rho )\text{ }d\mu =\int_\Omega [\rho ^d+\Lambda
(\rho ^c)]\text{ }d\mu =\int_\Omega \rho ^d\text{ }d\mu +\int_\Omega \Lambda
(\rho ^c)\text{ }d\mu =\int_\Omega \rho ^d\text{ }d\mu =1  \label{4.2}
\end{equation}
since in this case ${\cal L=D}^{\bot }$ and $1\in {\cal D}$, while $\Lambda
(\rho ^c)\in {\cal L}$, then: 
\begin{equation}
\int_\Omega \Lambda (\rho ^c)\text{ }d\mu =\int_\Omega \Lambda (\rho ^c).1%
\text{ }d\mu =\left\langle \Lambda (\rho ^c)\mid 1\right\rangle =0
\label{4.3}
\end{equation}

\noindent (for the same reason $\int_\Omega \rho ^cd\mu =0).$

Let us now consider $W_t=\Lambda U_t\Lambda ^{-1},$ where $\Lambda
^{-1}=\lambda ^{-1}(T)=\int_{{\Bbb R}}\frac 1{\lambda (s)}dE.$ Then, for any 
$\rho $ in the domain of $\Lambda ^{-1}$, we have:

\begin{eqnarray}
\left\| W_t\rho \right\| _{{\cal L}}^2 &=&\left\| \Lambda U_t\Lambda
^{-1}\rho \right\| _{{\cal L}}^2=\left\| (U_t^{\dagger }\Lambda U_t)\Lambda
^{-1}\rho \right\| _{{\cal L}}^2=  \nonumber \\
&=&\left\| \lambda (T+t)\lambda ^{-1}(T)\rho \right\| _{{\cal L}}^2=\int_{%
{\Bbb R}}\left[ \frac{\lambda (s+t)}{\lambda (s)}\right] ^2d\left\| E\rho
\right\| _{{\cal L}}^2
\end{eqnarray}
being the function $s\mapsto \left[ \frac{\lambda (s+t)}{\lambda (s)}\right]
^2$ non negative and bounded by the integrable function $1$. As $\frac{%
\lambda (s+t)}{\lambda (s)}$ goes monotonously to $0$ (see above eq. (2.6)),
from the Lebesque dominated convergence theorem [20] we have that $||W_t\rho
||\downarrow 0.$

\section{Synthesis of both formalisms.}

In this section, we will relate the formalism of the $\Lambda $ with the
formalism of a rigging with $\Phi _H.$

Let us consider the Koopman operator of the dynamic $U_t:{\cal L\rightarrow L%
}$, with certain $\Lambda :{\cal L\rightarrow L}$ We have proved that this
is equivalent to a rigging of ${\cal L}$ with a Hilbert space $\Phi _H$ with
an inner product: 
\begin{equation}
\left\langle \Lambda \sigma \mid \Lambda \rho \right\rangle _{\Phi
_H}=\left\langle \sigma \mid \rho \right\rangle _{{\cal L}}
\end{equation}

\noindent as well as with its antidual $\Phi _H^{\times }$. This rigging
defines the following operators:

1) $\Lambda :{\cal L}\rightarrow \Phi _H,$ namely function $\Lambda $, but
with a restricted range.

2) $\Lambda ^{\times }:\Phi _H^{\times }\rightarrow {\cal L}$, namely the
antitransposed former operator defined as: 
\begin{equation}
\left\langle \rho \mid \Lambda ^{\times }(F)\right\rangle _{{\cal L}%
}:=F\left( \Lambda ^{\dagger }\rho \right) =F\left( \Lambda \rho \right)
\label{5.2}
\end{equation}
where, for simplicity, only for ${\cal L}$ we have made the identification: $%
{\cal L=L}^{\times }.$

3) ${\cal R}:\Phi _H^{\times }\rightarrow \Phi _H,$ namely the Riesz
representation (already defined in eq. (3.1)), which is related with the
former operator by: 
\begin{equation}
{\cal R}=\Lambda \Lambda ^{\times }  \label{5.3}
\end{equation}
and its inverse ${\cal R}^{-1}:\Phi _H\rightarrow \Phi _H^{\times }.$

4) ${\cal R}|_{{\cal L}}=R\geq 0$, such that: 
\begin{equation}
R=\Lambda ^2\text{ , or which is equivalent, }\Lambda =\sqrt{R}  \label{5.4}
\end{equation}

5) $\Lambda ^{-1}$as an extension of the $\Lambda $ inverted operator, or
which is the same thing: 
\begin{equation}
\Lambda ^{-1}=\sqrt{R^{-1}}  \label{5.5}
\end{equation}

We also obtain some important operators combining the rigging with the
dynamics:

6) $W_t=\Lambda U_t\Lambda ^{-1},$ $t\in G^{+}$, namely the evolution
operator of the {\bf Markov semigroup} that we have already considered and
on which is based the $\Lambda $ formalism.

7) $\overline{U}_t:\Phi _H^{\times }\rightarrow \Phi _H^{\times },$ $t\in
G^{+},$ namely the extension of a {\bf semigroup} of $U_t$ to $\Phi
_H^{\times }$, defined by: 
\begin{equation}
\left( \overline{U_t}(F)\right) (\Lambda \rho )=F\left( U_{-t}\Lambda \rho
\right)  \label{5.6}
\end{equation}
which is the base of the rigging formalism.

8) $Y_t:{\cal L\rightarrow L},$ $t\in G^{+}$ defined as: 
\begin{equation}
Y_t=\Lambda ^{\times }\overline{U_t}\left( \Lambda ^{\times }\right) ^{-1}
\label{5.7}
\end{equation}

9) $V_t:\Phi _H^{\times }\rightarrow \Phi _H^{\times },$ $t\in G^{+}$
defined as: 
\begin{equation}
V_t=\left( \Lambda ^{\times }\right) ^{-1}U_t\Lambda ^{\times }  \label{5.8}
\end{equation}

10) $Z_t:\Phi _H\rightarrow \Phi _{H,}$ $t\in G^{+}$ defined as: 
\begin{equation}
Z_t={\cal R}\overline{U_t}{\cal R}^{-1}  \label{5.9}
\end{equation}

11) $X_t:\Phi _H^{\times }\rightarrow \Phi _H^{\times },$ $t\in G^{+}$
defined as: 
\begin{equation}
X_t={\cal R}^{-1}W_t{\cal R}  \label{5.10}
\end{equation}

The following propositions make clear the deep relation among all these
operators.

\noindent {\bf Theorem:}

\noindent {\it For any }$t\in G^{+},$ {\it we have:}

\noindent {\it i) }$V_t=X_t.$

\noindent {\it ii) }$V_t${\it \ defines a strong Markov process.}

\noindent {\it iii) }$V_t\neq \overline{U}_t.$

\noindent {\it iv) }$Y_t\neq U_t.$

\noindent {\it v) }$W_t\neq Z_t.$

\noindent {\it vi) }$Z_t${\it \ defines a dynamic which is equivalent to }$%
\overline{U}_t.$

\noindent {\bf Demonstration:}

Let $t\in G^{+,}F\in \Phi _H^{\times },$ and $\rho \in \Phi _H.$ Then we
have: 
\begin{equation}
X_t=(\Lambda \Lambda ^{\times })^{-1}W_t(\Lambda \Lambda ^{\times })=\left(
\Lambda ^{\times }\right) ^{-1}(\Lambda ^{-1}W_t\Lambda )\Lambda ^{\times
}=\left( \Lambda ^{\times }\right) ^{-1}U_t\Lambda ^{\times }=V_t
\label{5.11}
\end{equation}
and so (i) is demonstrated.

As Riesz representation is an isometric isomorphism, then $X_t$ in $\Phi
_H^{\times }$ is equivalent to $W_t$ in $\Phi _H.$ Now, we have just
demonstrated that $V_t=X_t$, so (ii) is also demonstrated.

In order to demonstrate (iii), it is enough to show that: 
\begin{equation}
\Lambda ^{\times }\overline{U_t}\neq U_t\Lambda ^{\times }  \label{5.12}
\end{equation}

Now, it is: 
\begin{equation}
\left\langle \rho \mid \Lambda ^{\times }\overline{U_t}(F)\right\rangle
=\left[ \overline{U_t}(F)\right] (\Lambda \rho )=F(U_{-t}\Lambda \rho )
\label{5.13}
\end{equation}
while: 
\begin{equation}
\left\langle \rho \mid U_t\Lambda ^{\times }(F)\right\rangle =\left\langle
U_{-t}\rho \mid \Lambda ^{\times }(F)\right\rangle =F(\Lambda U_{-t}\rho )
\label{5.14}
\end{equation}
As we know that $\Lambda $ do not commute with $U_{-t}$ (see (2.12)), it
turns out that the r.h.s. of the two last equations are not equal for all $F$
and for all $\rho $. Therefore eq. (5.12) is proved. So (iii) is
demonstrated.

As $(\Lambda ^{\times })^{-1}$ is a bijection , we have: 
\begin{equation}
Y_t=\Lambda ^{\times }\overline{U_t}\left( \Lambda ^{\times }\right)
^{-1}\neq \Lambda ^{\times }V_t\left( \Lambda ^{\times }\right) ^{-1}=U_t
\label{5.15}
\end{equation}
which proves (iv).

Finally, if we take into account (i), plus the relation that can be obtained
from eq. (5.10), and the fact that ${\cal R}$ is bijective: 
\begin{equation}
Z_t={\cal R}\overline{U_t}{\cal R}^{-1}\neq {\cal R}V_t{\cal R}^{-1}=W_t
\label{5.16}
\end{equation}

The last part is similar to the proof of (ii).

\section{Acknowledgement}

The author wishes to thank Professor Mario A.Castagnino and Dr. Roberto
Laura for many teachings, important dialogs and continuous encouragement,
and also to the ''unknown referee'', whose deep remarks helped me to
understand some mistakes.

\section{References}

\noindent [1] B.Misra, I.Prigogine, M.Courbage. Physica 98 A (1979) 1-26

\noindent [2] Idem Proc. Nat. Acad. of Sci. USA 76 (1979) 4768-72

\noindent 
[3] I.E.Antoniou, These, present\'{e}e dans l'Universit\'{e} Libre de
Bruxelles.

\noindent 
[4] S.Goldstein, B.Misra, M.Courbage. J.Stat.Phys. 25 (1981) 111-126

\noindent [5] I.E.Antoniou, I.Prigogine. Physica A 192 (1993) 443-464

\noindent [6] I.E.Antoniou, S.Tasaki. Int.Quantum Chemistry, 46 (1993),
425-474.

\noindent [7] Idem. Physica A, 190 (1992),303.

\noindent [8] I.Prigogine, ''Time, Chaos and the Laws of Nature'',
Conference, July 27 th, 1994.

\noindent [9] M.A.Castagnino, F.Gaioli, E.Gunzig. Found. of Cosmic Physic.
In press.

\noindent [10] A.B\"{o}hm, M.Gadella, ''Dirac Kets, Gamow vectors and
Gel'fand Triplets'',

\noindent Springer Verlag (1989).

\noindent [11] M.Cotlar, ''Equipaci\'{o}n con Espacios de Hilbert'', Cursos
y Seminarios de Matem\'{a}tica (N\'{u}mero 15, 1968), Univ. Buenos Aires.

\noindent [12] I.M.Gelfand, G.E.Chilov, ''Les Distributions'', vol. 2, Dunod
(1964)

\noindent [13] I.M.Gelfand, N.Y.Vilenkin, ''Les Distributions'', vol 4,
Dunod (1967)

\noindent [14] A.B\"{o}hm, ''The Rigged Hilbert Space and Quantum
Mechanics'', Lecture Notes in Physics, N\'{u}mero 78.

\noindent [15] I.Antoniou, Z.Suchanecki, in 'Nonlinear, deformed and
irreversible quantum systems'. H.D. Doebner et al editors, World Scientific
(1995)

\noindent [16] I.Antoniou, Z.Suchanecki, R.Laura, S.Tasaki, 'Intrinsic
irreversibility of quantum systems with diagonal singularity', Physica A,
241 (1997), 737-772.

\noindent [17] M.Reed, B.Simon, ''Methods of Modern Mathematical Physics'',
Acad.Press, vol. 1 (1979)

\noindent [18] M.Mackey,''Times Arrow: The Origins of Thermodynamic
Behaviour''Springer Verlag (1992)

\noindent [19] V.I.Arnold, A.Avez, ''Ergodic Problems of Classical
Mechanics'', Benjamin (1968)

\noindent [20] W.Rudin, ''Functional Analysis'', Mc. Graw Hill (1973)

\noindent [21] R.Thom, ''Structural Stability and Morphogenesis'',Benjamin
(1975)

\noindent [22] B.Misra, Proc.Natl.Acad.Sci.USA, vol 75, N4 (April 1978)
1627-31

\noindent [23] K.Maurin, ''General eigenfuction expansions and unitary
representations of topological groups'', Warzawa (1968)

\noindent [24] N.N.Bogolubov, A.A.Logunov and I.T.Todorov, ''Introduction to
Axiomatic Quantum Field Theory'', Benjamin (1975)

\noindent [25] R.Laura and A.R.Ord\'{o}\~{n}ez, ''Internal Time
Superoperator for Quantum Systems with Diagonal Singularity'', to appear in:
Inter.Jour.Theor.Phys. (1997)

\noindent [26]G.G.Emch, Commun. Math. Phys. {\bf 49}, 191-215, (1976)

\end{document}